\documentclass{ws-ijmpa}

\usepackage[super]{cite}
\usepackage{xcolor}
\usepackage[verbose,hypertexnames=false]{hyperref}
\hypersetup{colorlinks=false,allbordercolors=blue,pdfborderstyle={/S/U/W 1}}
\usepackage{placeins}   
\usepackage{graphicx}
\usepackage{tabularx}
\usepackage{slashed}
\usepackage{makecell}
\usepackage{booktabs}

\newcommand{\bma}{\left(\begin{matrix}}
\newcommand{\ema}{\end{matrix}\right)}

\newcommand{\mL}{\mathcal{L}}
\newcommand{\mM}{\mathcal{M}}

\begin{document}

\markboth{Bai,Hao,Guo}{Axion-like particle-meson production in semileptonic $\tau$ decays}

%
\catchline{}{}{}{}{}
%

\title{Axion-like particle-meson production in semileptonic $\tau$ decays}

\author{Yu-Xuan Bai,\, Jin Hao,\, Zhi-Hui Guo
}

\address{Department of Physics and Hebei Key Laboratory of Photophysics Research and Application,\\
Hebei Normal University,  Shijiazhuang 050024, China
}

%

\maketitle


\begin{abstract}
In this work we explore the semileptonic $\tau$ decays into the axion-like particle ($a$)-meson final states within chiral effective field theory. The next-to-leading-order mixing matrix for the $\pi^0$-$\eta$-$\eta'$-$a$ system with the linear isospin-breaking effects, is exploited and then implemented to calculate the hadronic form factors relevant to the $\tau$ decays. The resonance parameters entering the form factors are determined from fits to the experimental spectra of $\tau^- \to \pi^- \pi^0 \nu_\tau$, $\tau^- \to K_S \pi^- \nu_\tau$, and $\tau^- \to K^- \eta \nu_\tau$. We then focus on the predictions to  the branching ratios, invariant-mass distributions, and forward-backward asymmetries from the $\tau^-\to P a \nu_\tau$ processes, with $P=\pi^-$ and $K^-$. Our results provide a quantitative basis for future searches of the axion-like particle signals in semileptonic $\tau$ decays.
\end{abstract}


\ccode{PACS numbers: 13.35.Dx, 14.80.Va, 12.39.Fe}

\section{Introduction}

To address the long-standing strong CP problem, Peccei and Quinn proposed the PQ mechanism in Refs.~\cite{Peccei:1977hh,Peccei:1977ur}, which predicts a pseudo-Nambu-Goldstone boson (pNGB) beyond the Standard Model, viz. the axion~\cite{Weinberg:1977ma,Wilczek:1977pj}. In low-energy effective theories, the most characteristic axion interaction is the model-independent anomalous term $a G\tilde G/f_a$, where $G$ and $\tilde G$ correspond to the gluon field-strength tensor and its dual tensor, respectively, with $f_a$ the axion decay constant. The axion mass caused by this anomalous operator is given by $m_a=m_\pi F_\pi/(2f_a)$~\cite{Weinberg:1977ma}, where $m_\pi$ and $F_\pi$ correspond to the mass and decay constant of pion, implying that $m_a$ and $f_a$ are not independent. 
Meanwhile, axion-like particle (ALP), which together with axion will be denoted as $a$ throughout this work, presents a general extension of the axion, by introducing an additional source for the ALP mass. In this way, the mass and decay constant of the ALP can be independent, which leads to more rich phenomenological consequences~\cite{DiLuzio:2020wdo,Kim:2008hd,Graham:2015ouw,Irastorza:2018dyq,Choi:2020rgn}. The search for the axion and ALP signatures in particle physics, nuclear physics, astrophysics, cosmology, and even in condensed matter physics has become an active cross-disciplinary research frontier~\cite{DiLuzio:2020wdo,Kim:2008hd,Graham:2015ouw,Irastorza:2018dyq,Choi:2020rgn}. 

Since the $\tau$ lepton is heavy enough to decay into hadronic states, the semileptonic $\tau$ processes offer a useful platform for studying the low-energy QCD dynamics. This opportunity is becoming increasingly relevant at the future high-luminosity facilities, such as the super tau-charm facility (STCF)~\cite{Achasov:2023gey,Cheng:2025kpp}, since they are expected to provide much more precise measurements on the $\tau$ decays. In this context, the $\tau$ decay channels with two light pseudoscalars in the final states deserve renewed theoretical attention~\cite{Hao:2025pai,Li:2025zus,Cheng:2025kpp,Escribano:2016ntp,Aguilar:2024ybr}. In particular, the axion/ALP-meson channels have been much less systematically explored, which motivates the present analysis. For simplicity, we will collectively designate ALP and axion as ALP in later discussions.

Within the $U(3)$ chiral theory, the complete NLO $\pi^0$-$\eta$-$\eta'$-$a$ mixing matrix by including the linear isospin-breaking effect is determined in Refs.~\cite{Gao:2022xqz,Gao:2024vkw}. To describe the hadronic dynamics in the $\tau$-decay region, we combine this mixing matrix and construct the vector/scalar form factors for semileptonic $\tau$ decays into two-meson and ALP-meson final states within resonance chiral theory (R$\chi$T)~\cite{Ecker:1988te}, adopting the model-independent ALP interaction from the anomalous $a G\tilde G$ term. Both the Cabibbo-allowed decays $\tau^- \to (\pi P)^- \nu_\tau$ and the Cabibbo-suppressed ones $\tau^- \to (K P)^- \nu_\tau$ are considered, with $P=\pi,\eta,\eta',a$.
The various resonance parameters are fixed through a combined fit to the $\pi^-\pi^0$, $K_S\pi^-$, and $K^-\eta$ invariant-mass spectra measured in $\tau^- \to \pi^- \pi^0 \nu_\tau$, $\tau^- \to K_S \pi^- \nu_\tau$, and $\tau^- \to K^- \eta \nu_\tau$ decays~\cite{Hao:2025pai}. With these inputs, we obtain predictions for several unmeasured channels, including $\tau^- \to K^- \eta' \nu_\tau$, $\tau^- \to \pi^- \eta(\eta') \nu_\tau$, and especially $\tau^- \to (\pi/K)^- a \nu_\tau$, together with their branching ratios, invariant-mass spectra, and corresponding forward-backward asymmetries. These results can provide useful phenomenological guidelines for future studies of ALP productions in the semileptonic $\tau$ decays.

\section{Chiral Lagrangian and $\pi^0$-$\eta$-$\eta'$-$a$ mixing formalism}

Throughout this work, we take the model-independent anomalous ALP-gluon interaction term $a\,G\tilde G/f_a$ and ignore any direct ALP-lepton or ALP-quark interactions. The corresponding ALP effective Lagrangian can be written as
\begin{eqnarray}\label{eq.lagag}
\mathcal{L}_a^{G}= \frac{1}{2}\partial_\mu a \partial^\mu a + \frac{a}{f_a} \frac{\alpha_s}{8\pi}G^i_{\mu\nu}\tilde{G}^{i,\mu\nu}  - \frac{1}{2} m_{a,0}^{2}  \, a^2 \,,
\end{eqnarray}
where $G^i_{\mu\nu}$ is the gluon field-strength tensor, $\tilde{G}^{i,\mu\nu}$ is its dual with $i$ the color index, and $m_{a,0}$ is the bare ALP mass. For the QCD axion, the bare mass $m_{a,0}$ is set to zero. In the $U(3)$ chiral theory, we do not have to explicitly eliminate the $a\,G\tilde G/f_a$ term by performing an axial rotation of the quark fields, as usually done in the $SU(2)$ and $SU(3)$ case~\cite{GrillidiCortona:2015jxo}. Instead, we incorporate the ALP field into the $U_A(1)$ anomalous term and the relevant leading-order (LO) Lagrangian is 
\begin{eqnarray}\label{eq.laglo}
\mL^{\rm LO}= \frac{ F^2}{4}\langle u_\mu u^\mu \rangle+
\frac{F^2}{4}\langle \chi_+ \rangle
+ \frac{F^2}{12}M_0^2 X^2 \,,
\end{eqnarray}
where the mass-square term $M_0^2$ of the singlet $\eta_0$ is proportional to the topological susceptibility, behaving as $O(1/N_C)$. In the counting rule of $\delta$ expansion~\cite{Kaiser:2000gs}, $M_0^2$ together with the momentum squared $p^2$ and light-quark mass $m_q$, are all counted as $O(\delta)$. The ALP field is introduced into the chiral Lagrangian through $X= \log{(\det U)} - i a/f_a$\,. For details of the definitions of other chiral building terms, see Ref.~\cite{Gao:2024vkw} and references therein.

In the $\tau$-decay energy region, hadronic resonances play an essential role, so the LO chiral Lagrangian alone is insufficient to describe the energy spectra. We employ the R$\chi$T to include interactions with both vector and scalar resonance multiplets~\cite{Ecker:1988te}
\begin{equation}\label{eq.lagvs}
\begin{aligned}
\mathcal{L}_{V,S}^{U(3)} = \frac{F_V}{2\sqrt{2}} \langle V_{\mu\nu} f_+^{\mu\nu} \rangle + \frac{iG_V}{\sqrt{2}} \langle V_{\mu\nu} u^\mu u^\nu \rangle + c_d \langle S u_\mu u^\mu \rangle + c_m \langle S \chi_+
\rangle \,,
\end{aligned}
\end{equation}
where $V_{\mu\nu}$ represents the vector resonant nonet state in antisymmetric tensor form, and $S$ stands for the scalar resonance fields. In the phenomenological analysis of this work, we apply the large $N_C$ relationship to the vector couplings $(F_V,G_V)$ and scalar couplings $(c_d,c_m)$ to reduce the number of free parameters~\cite{Hao:2025pai}. In addition, we also include the excited multiplets of vector and scalar resonances, whose parameters are denoted by $(F_V^{',''},G_V^{',''})$ and $(c_d',c_m')$. Two additional NLO $U(3)$ operators with coefficients $\Lambda_1$ and $\Lambda_2$ are also included
\begin{align}\label{eq.lambda}
\mL^{U(3)}_{\Lambda_i}=-\frac{F^2\, \Lambda_1}{12}   D^\mu X D_\mu X  -\frac{F^2\, \Lambda_2}{12} X \langle \chi_- \rangle\,,
\end{align}
which can not be generated from the resonance operators in Eq.~\eqref{eq.lagvs}. 

At LO in the $\delta$ counting scheme, only the mass term of $\pi^0,\eta,\eta'$ and $a$ is mixed, while at NLO, both the kinetic and mass terms are mixed. We can calculate the mixing of $\pi^0,\eta,\eta'$ and $a$ order by order in the $\delta$ counting. We denote the bare fields by $\pi^0$, $\eta_8$, $\eta_0$ and $a$, and the LO diagonalized fields by $\overline{\pi}^0$, $\overline{\eta}$, $\overline{\eta}'$ and $\overline{a}$. By incorporating the linear correction of isospin-breaking, the relationship between the LO diagonalized fields and the bare fields can be obtained through the mass mixing term of the LO Lagrangian Eq.~\eqref{eq.laglo}, which is given by
\begin{equation}\label{eq.lomat}
\begin{aligned}
\left( \begin{array}{c}
\overline{\pi}^0 \\  \overline{\eta} \\ \overline{\eta}' \\ \overline{a} 
\end{array} \right) \,&=\text{M}^\text{LO}\left( \begin{array}{c}
\pi^0 \\  \eta_8 \\ \eta_0 \\ a
\end{array} \right) \,=\,
 \left( \begin{array}{cccc}
1  & -(c_\theta v_{12} + s_\theta v_{13})  & s_\theta v_{12} - c_\theta v_{13} & -v_{14} \\ 
v_{12}  & c_\theta - s_\theta v_{23}  & -(s_\theta + c_\theta v_{23})    & -v_{24}  \\ 
v_{13}  & c_\theta v_{23} +s_\theta   &  c_\theta -s_\theta v_{23}     & -v_{34} \\ 
v_{41}   & c_\theta v_{42} +s_\theta v_{43}   & c_\theta v_{43} -s_\theta v_{42}  & 1+ v_{44}
\end{array} \right)\,
\left( \begin{array}{c}
\pi^0 \\  \eta_8 \\ \eta_0 \\ a
\end{array} \right) \,, 
\end{aligned}
\end{equation}
where $\mathrm{M}^{\mathrm{LO}}$ represents the LO four-particle mixing matrix, and $c_\theta=\cos\theta$ and $s_\theta=\sin\theta$, being $\theta$ the LO $\eta_8$-$\eta_0$ mixing angle~\cite{Guo:2015xva}.
The matrix element $v_{ij}$ can be obtained from the mass mixing term at LO of the Lagrangian Eq.~\eqref{eq.laglo}. Here we only provide the expressions of two specific terms $v_{14}$ and $v_{41}$ in Eq.~\eqref{eq.lomat} as examples
\begin{eqnarray}\label{eq.v12}
&v_{41}= -\frac{ M_0^2 \epsilon}{6  (m_a^2 - m_{\overline{\pi}}^2)} \frac{F}{f_a}\bigg[-\frac{(\sqrt{2} c_\theta - 
       2 s_\theta) s_\theta}{ m_a^2 - m_{\mathring{\overline{\eta}}}^2} + \frac{c_\theta (2 c_\theta + \sqrt{2} s_\theta)}{ m_a^2 - m_{\mathring{\overline{\eta}}'}^2}\bigg]\,,\\
&v_{14}= -\frac{ M_0^2 \epsilon}{6 (m_a^2 - m_{\overline{\pi}}^2)}\frac{F}{f_a}\bigg[-\frac{(\sqrt{2} c_\theta - 
       2 s_\theta) s_\theta}{ m_{\overline{\pi}}^2 - m_{\mathring{\overline{\eta}}}^2} + \frac{c_\theta (2 c_\theta + \sqrt{2} s_\theta)}{ m_{\overline{\pi}}^2 - m_{\mathring{\overline{\eta}}'}^2}\bigg] \,,
\end{eqnarray}
where $\epsilon\equiv B(m_u-m_d)$ is the correction term for the isospin-breaking effect from QCD. In the phenomenological study, we estimate its value using $\epsilon = m_{K^{+}}^2 - m_{K^0}^2 - (m_{\pi^{+}}^2 - m_{\pi^0}^2)$. Other expressions of the remaining matrix elements $v_{ij}$ in Eq.~\eqref{eq.lomat} can be found in Refs.~\cite{Gao:2022xqz,Gao:2024vkw}. 
At NLO, the LO diagonalized fields $\overline{\pi}^0$, $\overline{\eta}$, $\overline{\eta}'$ and $\overline{a}$ will be mixed again. The relevant NLO operators are the ones accompanied by the low-energy constants (LECs) $L_5$ and $L_8$, together with the terms involving $\Lambda_1$ and $\Lambda_2$~\cite{Gao:2022xqz}. In the framework of R$\chi$T, the LECs are assumed to be saturated by the resonance states, leading to $L_5=c_mc_d/M_S^2+c_m'c_d'/M_{S'}^2$ and $L_8=c_m^2/(2M_S^2)+c_m'^2/(2M_{S'}^2)$. For the LECs $\Lambda_1$ and $\Lambda_2$, we will take their values from Ref.~\cite{Gao:2022xqz}. The relation between the bare fields $(\pi^0,\eta_8,\eta_0,a)$ and the NLO diagonalized fields $(\hat{\pi}^0,\hat{\eta},\hat{\eta}',\hat a)$ can then be established through the following matrix transformation
\begin{equation}\label{NLOmixing}
{\small
\begin{aligned}
    \left( \begin{array}{c}
\hat{\pi}^0 \\   \hat{\eta}  \\  \hat{\eta}' \\ \hat{a} 
\end{array} \right)
&
\,=\left\{\text{M}^\text{LO} + \left(\begin{array}{cccc}
 z_{11}  &  c_\theta z_{12} + s_\theta z_{13}  &  c_\theta z_{13}-s_\theta z_{12}  & z_{14} \\ 
z_{21}  & c_\theta z_{22} + s_\theta z_{23}  &  c_\theta z_{23}-s_\theta z_{22}     & z_{24}  \\ 
z_{31}  & c_\theta z_{32} +s_\theta z_{33}  & c_\theta z_{33}  -s_\theta z_{32}   & z_{34} \\ 
z_{41}   & c_\theta z_{42} +s_\theta z_{43}   & c_\theta z_{43} -s_\theta z_{42} & z_{44} 
\end{array} \right)\right\} \left( \begin{array}{c}
\pi^0 \\  \eta_8 \\  \eta_0 \\ a
\end{array} \right) \,,
\end{aligned}
}
\end{equation}
where $\text{M}^{\text{LO}}$ is already given in Eq.~\eqref{eq.lomat} and the explicit expressions for the NLO corrections encoded in the matrix elements $z_{ij}$ can be found in Ref.~\cite{Gao:2024vkw} and will not be repeated here. The mixing matrix in Eq.~\eqref{NLOmixing} allows us to calculate the ALP-meson form factors from the chiral Lagrangians in Eqs.~\eqref{eq.laglo}-\eqref{eq.lambda}. 

\section{$\tau$ decay amplitudes and form-factor calculations}

In the Standard Model, the amplitude of the decay process $\tau(p_\tau) \to P_1(p_1)\, P_2(p_2)\, \nu_\tau(p_\nu)$ can be written as 
\begin{equation}\label{eq.defmt}
 \mM= \frac{G_FV_{uD}}{\sqrt2}\, L_\mu H^\mu\,, \quad (D=d,s)\,,
\end{equation}
where $G_F$ is the Fermi constant, $V_{ud/us}$ denote the CKM matrix elements, the leptonic current is given by $L_\mu=\bar{u}(p_\nu)\gamma_\mu(1-\gamma_5)u(p_\tau)$ and the hadronic part can be parameterized as~\cite{Hao:2025pai} 
\begin{equation}\label{eq.defhmu}
H^\mu= \left[ (p_{2} - p_{1})^\mu - \frac{\Delta_{P_2P_1}}{s} q^\mu \right] F^{P_1P_2}_{+}(s) +  \frac{\Delta_{Du}^{\text{Phy}}}{s} q^\mu F^{P_1P_2}_{0}(s)\,,
\end{equation}
with $\Delta_{P_2P_1}=m_{P_2}^2-m_{P_1}^2\,,  q_\mu =(p_1+p_2)_\mu\,,  s=q^\mu q_\mu\,,$
$\Delta_{du}^{\text{Phy}}= m_{K^0}^2 -m_{K^{+}}^2 - (m_{\pi^0}^2- m_{\pi^{+}}^2 )\,,$ and $\Delta_{su}^{\text{Phy}}=m_K^2-m_\pi^2\,.$
$F^{P_1P_2}_{+}(s)$ and $F^{P_1P_2}_{0}(s)$ denote the vector and scalar form factors, respectively.

Starting from the amplitude in Eq.~\eqref{eq.defmt}, one obtains the differential decay width for the $\tau^- \to (P_1P_2)^- \nu_\tau$ process 
\begin{equation}\label{dGammapi}
\begin{aligned}
\frac{d\Gamma_{\tau \to P_1 P_2 \nu_\tau}}{d\sqrt{s}} =& \frac{G_F^2 M_\tau^3}{48 \pi^3 s}\, S_\text{EW}\, \left| V_{uD} \right|^2 \left( 1 - \frac{s}{M_\tau^2} \right)^2 \\& 
\times\left\{
\left( 1 + \frac{2s}{M_\tau^2} \right) q_{P_1P_2}^3(s) \left| F_{+}^{P_1P_2}(s) \right|^2
+ \frac{3 {\Delta_{Du}^{\text{Phy}}}^2}{4s} q_{P_1P_2}(s) \left| F_{0}^{P_1P_2}(s) \right|^2
\right\}\,,
\end{aligned}
\end{equation}
where $q_{P_1P_2}(s) = \frac{\sqrt{s^2 - 2s \Sigma_{P_1P_2} + \Delta_{P_1P_2}^2}}{2\sqrt{s}}$, $\Sigma_{P_1P_2} = m_{P_1}^2 + m_{P_2}^2$, 
and $\sqrt{s}$ denotes the invariant mass of the $P_1P_2$ pair in their center-of-mass (CM) frame. Throughout this work, the short-distance electroweak correction factor is taken to be $S_{\text{EW}}=1.0201$~\cite{Erler:2002mv}. In addition to the differential decay width in Eq.~\eqref{dGammapi}, we will also consider the forward-backward asymmetry $A_{FB}$, which can be written as~\cite{Hao:2025pai} 
\begin{equation}\label{eq.afb}
\begin{aligned}
A_{FB}(s) 
=& \frac{ \Delta_{Du}^{\text{Phy}} \, q_{P_1P_2}(s) \, \Re \left[  F_+^{P_1P_2}(s)  \, F_0^{P_1P_2*}(s) \right]}{\frac{2\sqrt{s}}{3}\left( 1 + \frac{2s}{M_\tau^2} \right) q_{P_1P_2}^2(s) \left| F_{+}^{P_1P_2}(s) \right|^2
+ \frac{ {\Delta_{Du}^{\text{Phy}}}^2}{2\sqrt{s}}  \left| F_{0}^{P_1P_2}(s) \right|^2
}\,.
\end{aligned}
\end{equation}
This observable is highly sensitive to the vector-scalar interference. Such an interference effect does not appear in the differential decay width in Eq.~\eqref{dGammapi}.

Next, the relevant hadronic form factors are evaluated within R$\chi$T. We include three multiplets of the vector resonances---the ground-state multiplet together with two excited ones, and two multiplets of the scalar resonances. The four-state mixing formula in Eq.~\eqref{NLOmixing} are incorporated throughout to calculate the relevant form factors. An important point is that the NLO part of the matrix elements (i.e., the terms involving $z_{ij}$) depends on the LECs $L_5$ and $L_8$~\cite{Gao:2024vkw}. To ensure a consistent treatment, the resonance-saturation estimation for $L_5$ and $L_8$ is adopted. 

As an illustrative example, we give the explicit expressions of the vector/scalar form factors for the channel $\tau^- \to \pi^- a \nu_\tau$ below in Eqs.~\eqref{VFFpia}--\eqref{eq.SFFpia}. For the vector case, in addition to the ground-state $\rho(770)$, two excited states, $\rho(1450)$ and $\rho(1700)$ (denoted by $\rho'$ and $\rho''$, respectively) are included in the form factor
\begin{equation}\label{VFFpia}
\begin{aligned}
F^{\pi^- a}_+(s)=&-\sqrt{2}\bigg\{\frac{ v_{41}}{F^2}G_{\rm{LO}+\rm{\rho\,Ex}}(s)+ y_{14} + v_{12}\, y_{24}^{(0)} + v_{13}\, y_{34}^{(0)} \bigg\} \,, 
\end{aligned}
\end{equation}
with 
\begin{equation}\label{eq.grhoex}
\small{
\begin{aligned}
G_{\rm{LO}+\rm{\rho\,Ex}}(s)  = &  \frac{ G_V F_V s+F^2(M_\rho^2-s)}{M_\rho^2 - s -iM_\rho\Gamma_\rho(s)}  -  \frac{G_V' F_V' s}{M_{\rho'}^2 - s -iM_{\rho'}\Gamma_{\rho'}(s)}-\frac{ G_V'' F_V'' s}{M_{\rho''}^2 - s -iM_{\rho''}\Gamma_{\rho''}(s)} \,. 
\end{aligned}
}
\end{equation}
For the scalar case, we include both the ground-state scalar resonance $a_0(980)$ (denoted by $a_0$) and the excited scalar resonance $a_0(1450)$ (denoted by $a_0'$)  

\begin{equation}\label{eq.SFFpia}
\small{
\begin{aligned}
F^{\pi^- a}_0(s)=&\frac{ (\sqrt{2} c_\theta -2 s_\theta )}{\sqrt{3}} v_{24}^{(0)} + \frac{ (2 c_\theta  +\sqrt{2} s_\theta )}{\sqrt{3}}v_{34}^{(0)} -\frac{2}{\sqrt{3}} \left( s_\theta \, v_{24}^{(0)} - c_\theta \, v_{34}^{(0)} + \frac{F}{\sqrt{6}f_a}\right) (\Lambda_2-\frac{1}{2} \Lambda_1)\\ & + \frac{ (\sqrt{2} c_\theta -2 s_\theta )}{\sqrt{3}} y_{24}^{(0)}  + \frac{(2 c_\theta  +\sqrt{2} s_\theta )}{\sqrt{3}} y_{34}^{(0)}+\frac{ 4(\sqrt{2} c_\theta -2 s_\theta )v_{24}^{(0)}+4(2 c_\theta  +\sqrt{2} s_\theta )v_{34}^{(0)}}{\sqrt{3}F^2} \\ &\bigg\{\bigg[\frac{  2c_m^2\, m_\pi^2 +  c_m c_d \left( s - m_a^2- m_\pi^2 \right) }{   M_{a_0}^2 - s -i M_{a_0} \Gamma_{a_0}(s)}  + \frac{2c_m(c_d-c_m)\,(2 m_K^2-m_\pi^2)}{ M_{S}^2 }  \bigg]\\ & + \bigg[ c_{m,d},M_{a_0},\Gamma_{a_0},M_{S} \to c_{m,d}',M_{a_0'},\Gamma_{a_0'},M_{S'}  \bigg]\bigg\}\,.
\end{aligned}
}
\end{equation}
$\Gamma_{\rho}(s)$, $\Gamma_{\rho',\rho''}(s)$, and $\Gamma_{a_0,a_0'}(s)$ are the energy-dependent widths, whose explicit expressions can be found in Ref.~\cite{Hao:2025pai}. At present, the available $\tau$-decay data are insufficient to constrain the masses and widths of the scalar resonances $a_0(980)$ and $a_0(1450)$. Therefore, these parameters are not treated as free parameters in the fits. Instead, they are fixed using the pole positions on the complex energy plane quoted by the PDG~\cite{ParticleDataGroup:2024cfk}. The same strategy can also be applied to determine the resonance parameters associated with the scalar resonances $K_0^\ast(700)$ and $K_0^\ast(1430)$ in strangeness-changing processes.
The explicit expressions for the form factors for the final states $\pi\pi$, $\pi\eta^{(')}$, $K\pi$, $K\eta^{(')}$, and $Ka$ for $\tau$ decays can be found in Ref.~\cite{Hao:2025pai} and are not reproduced here.

\section{Phenomenological studies of ALP-meson production from $\tau$ decays}

A joint fit to three sets of experimental data in the $\tau$ decays has been performed to determine the relevant resonance parameters in Ref.~\cite{Hao:2025pai}. The data sets used are the $\pi^-\pi^0$ vector form factor modulus squared $|F^{\pi^-\pi^0}_+(s)|^2$ measured in $\tau^- \to \pi^-\pi^0\nu_\tau$~\cite{Belle:2008xpe} and the invariant-mass spectra of the hadronic systems in the decay channels $\tau^- \to  K_S\pi^-\nu_\tau$ and $\tau^- \to  K^-\eta\nu_\tau$ from Refs.~\cite{Belle:2007goc} and~\cite{Belle:2008jjb}. The parameter setups and the outputs from the joint fit, together with the comparisons of the theoretical curves and experimental data, are described in detail in Ref.~\cite{Hao:2025pai} and interested readers can find all the relevant information there.

Here we focus on the discussions of the various quantities related to the ALP-meson productions in the $\tau$ decays. The predicted branching ratios of $\tau$ decays into ALP-meson final states are collected in Table~\ref{BR}, where we choose three different scenarios for illustrating purposes, i.e., by taking the QCD axion ($m_a=0$) and two specific ALP masses at $m_a=0.1$ and 0.3~GeV. According to the numbers in the Table~\ref{BR}, one can conclude that the ALP-meson productions in $\tau$ decays are dominated by vector form factors, while the scalar form factors only play minor roles. 
Furthermore, we find that the inclusion of hadronic resonances can increase the branching ratios of the ALP productions by roughly an order of magnitude compared with the LO chiral result. Taking $m_a=0.1/0.3~\mathrm{GeV}$ as examples, the branching ratios of $\tau^- \to \pi^- a\nu_\tau$ obtained with the full hadronic amplitude are enhanced by factors of about $7.6/6.8$ relative to the LO approximation, whereas for $\tau^-\to K^- a\nu_\tau$ the enhancement is even more pronounced, by factors of about $19.5/11.9$. This demonstrates that to properly incorporate the effects of the hadronic resonances in the ALP amplitudes is clearly helpful to provide reliable theoretical inputs when estimating the ALP parameter bounds. Recent progresses along this research direction have been made for various ALP reactions, such as the $a\pi\to\pi\pi$ and $a K\to \pi K$ scattering~\cite{Wang:2023xny,Wang:2025wlu}, the $\eta\to\pi\pi a$ decay~\cite{Wang:2024tre}, the ALP photoproduction $\gamma N \to a N$~\cite{Cao:2024cym} and the $\ell N \to \ell N a$ processes~\cite{Guo:2025icf}. 

\begin{table}[htbp]
\centering
\renewcommand{\arraystretch}{1.2}
\begin{tabular}{cccc}
\toprule
Channel & Total & Vector & Scalar    \\
\midrule
$m_a=0$ & & & \\
$f_a^2 {\rm BR}_{\tau^- \to \pi^- a\nu_\tau}\,(10^{5}\ \mathrm{GeV}^2)$
& 4.44  
& 4.44  
& 0.01 
\\ 
$f_a^2 {\rm BR}_{\tau^- \to K^- a\nu_\tau}\,(10^{5}\ \mathrm{GeV}^2)$
& 1.48  
& 1.43  
& 0.05 
\\ \hline
$m_a=0.1$~GeV &  & & \\
$f_a^2 {\rm BR}_{\tau^- \to \pi^- a\nu_\tau}\,(10^{5}\ \mathrm{GeV}^2)$
& $21.8$
& $21.8$
& 0.004 
\\ 
$f_a^2 {\rm BR}_{\tau^- \to K^- a\nu_\tau}\,(10^{5}\ \mathrm{GeV}^2)$
& $2.35$
& $2.29$
& 0.06  
\\ \hline
$m_a=0.3$~GeV &  & & \\
$f_a^2 {\rm BR}_{\tau^- \to \pi^- a\nu_\tau}\,(10^{5}\ \mathrm{GeV}^2)$
& $0.34$
& $0.34$
& $0.003$
\\ 
$f_a^2 {\rm BR}_{\tau^- \to K^- a\nu_\tau}\,(10^{5}\ \mathrm{GeV}^2)$
& $0.74$
& $0.73$
& $0.01$
\\
\bottomrule
\end{tabular}
\caption{Theoretical predictions for the branching ratios (BRs) of the $\tau\to P a \nu_\tau$ channels with different values of $m_a$. The last two columns give the individual contributions from the vector and scalar form factors. }
\label{BR}
\end{table}

\begin{figure}[!htbp]
  \centering
  \includegraphics[width=0.95\linewidth]{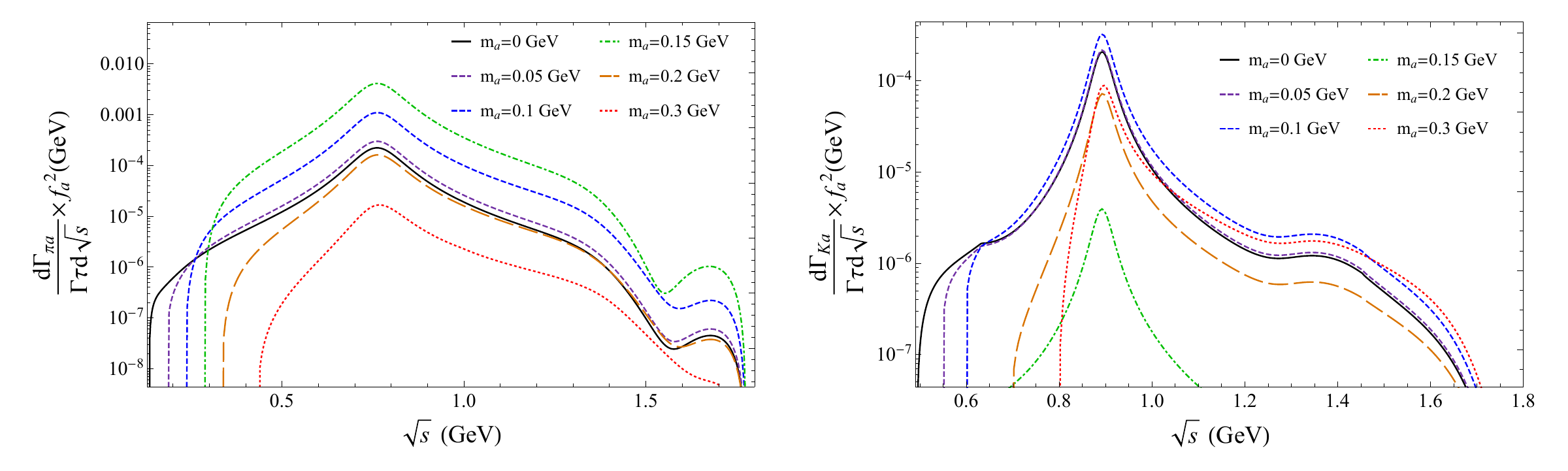}
  \caption{Differential decay widths for $\tau^- \to \pi^- a \nu_\tau$ (left) and $\tau^- \to K^- a \nu_\tau$ (right) as functions of $\sqrt{s}$, the invariant mass of the ALP-meson system. Results are shown for ALP masses with $m_a=0.05,\,0.1,\,0.15,\,0.2$, and $0.3$~GeV, the QCD-axion benchmark with $m_a=0$ is also displayed for comparison.}
  \label{piamaandKama}
  \end{figure}

We further extend our discussion to the differential decay widths as functions of the invariant masses of the $(\pi/K)^- a$ systems, as illustrated in Fig.~\ref{piamaandKama}. The results with several representative ALP masses with $m_a=0.05,\,0.1,\,0.15,\,0.2$ and $0.3~\mathrm{GeV}$ are shown, and the curves for $m_a=0$, corresponding to the QCD-axion benchmark scenario, are also included for comparison. 

\begin{figure}[!htbp]
\includegraphics[width=0.95\linewidth]{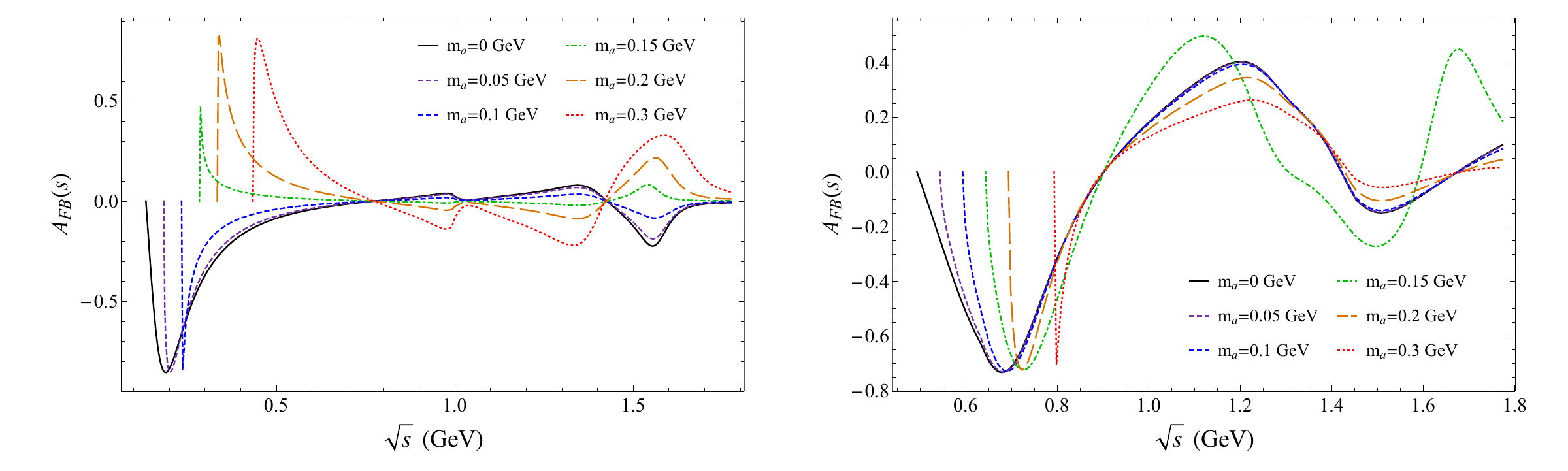}
  \caption{ Forward-backward asymmetries $A_{FB}$ for $\tau^- \to \pi^- a \nu_\tau$ (left) and $\tau^- \to K^- a \nu_\tau$ (right) as functions of $\sqrt{s}$, the invariant mass of the ALP-meson system. Results are shown for ALP masses with $m_a=0.05,\,0.1,\,0.15,\,0.2$, and $0.3~\mathrm{GeV}$, the QCD-axion benchmark with $m_a=0$ is also displayed for comparison.}
  \label{AFBpiamaandKama}
\end{figure}

Finally, we provide the theoretical predictions to the forward-backward asymmetries $A_{FB}$, defined in Eq.~\eqref{eq.afb}. Here we focus on the distributions of $A_{FB}$ as functions of the invariant mass of the ALP-meson system for $\tau^- \to \pi^- a \nu_\tau$ and $\tau^- \to K^- a \nu_\tau$ channels at different ALP masses, as displayed in Fig.~\ref{AFBpiamaandKama}; the QCD axion case is included for comparison as well.  

\section{Summary and outlook}

By relying on the model-independent anomalous ALP interaction $a G\tilde G/f_a$, we have studied the semileptonic $\tau$ decays into the ALP-meson final states in a unified framework within resonance chiral theory. The unknown hadronic resonance parameters entering the ALP-meson production amplitudes are determined through a combined fit to the experimental invariant-mass spectra measured by the Belle Collaboration in $\tau^- \to \pi^- \pi^0 \nu_\tau$, $\tau^- \to K_S \pi^- \nu_\tau$, and $\tau^- \to K^- \eta \nu_\tau$ decays. We then explore the theoretical predictions to the branching ratios, invariant-mass distributions and the forward-backward asymmetries for the $\tau^- \to \pi^-/K^- a \nu_\tau$ processes, by taking several different values for the ALP masses. 
We find that hadronic resonance effects can greatly enhance the production rates for the aforementioned processes, compared with the LO chiral result. These results provide useful phenomenological benchmarks for future searches of the ALP in semileptonic $\tau$ decays at future high-luminosity facilities, such as the STCF. 

\section*{Acknowledgements}

This work is partially supported by the National Natural Science Foundation of China (NSFC) under Grants No.~12475078, No.~12150013, No.~11975090, and also by the Science Foundation of Hebei Normal University with Contract No.~L2023B09.

\bibliography{refs}
\bibliographystyle{ws-ijmpa}

\end{document}